\documentclass{PoS}

\PoS{PoS(LAT2005)113}

\usepackage{amsmath}
\usepackage{epsfig}

\title{An Efficient Cluster Algorithm for CP(N-1) Models }

\ShortTitle{Efficient Cluster Algorithms for CP(N-1) Models }

\author{Michele Pepe, \speaker{St\'ephane Riederer}\thanks{This work
    was supported in part by the Swiss Nationalfonds for Scientific
    Research.}, Uwe-Jens Wiese\\
        Bern University, Switzerland\\
        E-mail: \email{pepe@itp.unibe.ch},
        \email{riederer@itp.unibe.ch}, \email{wiese@itp.unibe.ch}}

\author{Bernard B. Beard\\
        Christian Brothers University, Memphis, USA\\
        E-mail: \email{bbbeard@cbu.edu}}

\abstract{We construct an efficient cluster algorithm for ferromagnetic
  $SU(N)$-symmetric quantum spin systems. Such systems
provide a new regularization for $CP(N-1)$ models in the framework of
$D$-theory, which is an alternative non-perturbative approach to
quantum field theory formulated in terms of discrete quantum
variables instead of classical fields. Despite several attempts, no
efficient cluster algorithm has been constructed for $CP(N-1)$ models
in the standard formulation of lattice field theory. In
fact, there is even a no-go theorem that prevents the construction of
an efficient Wolff-type embedding algorithm. We present various
simulations for different correlation lengths, couplings and lattice
sizes. We have simulated correlation lengths up to 250 lattice
spacings on lattices as large as 640$\times$640 and we detect no evidence for
critical slowing down.}

\FullConference{XXIIIrd International Symposium on Lattice Field Theory\\
                 25-30 July 2005\\
                 Trinity College, Dublin, Ireland}

\begin{document}

\section{Standard Formulation of $CP(N-1)$ Models}
The manifold $CP(N-1) = SU(N)/U(N-1)$ is a $(2N-2)$-dimensional coset space 
relevant in the context of the spontaneous breakdown of an $SU(N)$ symmetry to 
a $U(N-1)$ subgroup. In particular, in more than two space-time dimensions 
($d > 2$) the corresponding Goldstone bosons are described by $N \times N$ 
matrix-valued fields $P(x) \in CP(N-1)$ which obey
\begin{equation}
P(x)^2 = P(x), \ P(x)^\dagger = P(x), \ \mbox{Tr} P(x) = 1.
\end{equation}
For $d = 2$ the Hohenberg-Mermin-Wagner-Coleman theorem implies that the 
$SU(N)$ symmetry cannot break spontaneously. Correspondingly, similar
to 4-dimensional
non-Abelian gauge theories, the fields $P(x)$ develop a mass-gap 
nonperturbatively. Motivated by these observations, D'Adda, Di Vecchia, and
L\"uscher~\cite{DAd78} introduced $CP(N-1)$ models as interesting toy models
for QCD. The corresponding Euclidean action is given by
\begin{equation}
\label{CPNaction}
S[P] = \int d^2x \ \frac{1}{g^2} \mbox{Tr}[\partial_\mu P \partial_\mu P],
\end{equation}
where $g^2$ is the dimensionless coupling constant.
Note that this action is invariant under global $\Omega \in SU(N)$ 
transformations 
\begin{equation}
P(x)' = \Omega P(x) \Omega^\dagger,
\end{equation}
and under charge conjugation $C$ which acts as $^CP(x) = P(x)^*$.

\section{D-Theory Formulation of $CP(N-1)$ Models}

In this section we describe an alternative formulation of field theory in which
the $2$-dimensional $CP(N-1)$ model emerges from the dimensional reduction of 
discrete variables --- in this case $SU(N)$ quantum spins in $(2+1)$
space-time 
dimensions. The dimensional reduction of discrete variables is the 
key ingredient of D-theory, which provides an alternative nonperturbative
regularization of field theory. In D-theory we start from a ferromagnetic 
system of $SU(N)$ quantum spins located at the sites $x$ of a $2$-dimensional 
periodic square lattice. The $SU(N)$ spins are represented by
Hermitean operators $T_x^a = \frac{1}{2} \lambda_x^a$ (Gell-Mann matrices for 
the triplet representation of $SU(3)$) that generate the group $SU(N)$ and thus
obey
\begin{equation}
[T_x^a,T_y^b] = i \delta_{xy} f_{abc} T_x^c, \ 
\mbox{Tr}(T_x^a T_y^b) = \frac{1}{2} \delta_{xy} \delta_{ab}.
\end{equation}
In principle, these generators can be taken in any irreducible representation 
of $SU(N)$. However, as we will see later, not all representations lead to 
spontaneous symmetry breaking from $SU(N)$ to $U(N-1)$ and thus to $CP(N-1)$ 
models. The Hamilton operator for an $SU(N)$ ferromagnet takes the form
\begin{equation}
H = - J \sum_{x,i} T_x^a T_{x+\hat i}^a,
\end{equation}
where $J>0$ is the exchange coupling. By construction, the Hamilton operator is
invariant under the global $SU(N)$ symmetry, i.e.\ it commutes with the total 
spin given by 
\begin{equation}
T^a = \sum_x T_x^a.
\end{equation}

The Hamiltonian $H$ describes the evolution of the quantum spin system in an 
extra dimension of finite extent $\beta$. In D-theory this extra dimension is 
not the Euclidean time of the target theory, which is part of the 
$2$-dimensional lattice. Instead, it is an additional compactified dimension 
which ultimately disappears via dimensional reduction. The quantum partition 
function
\begin{equation}
Z = \mbox{Tr} \exp(- \beta H)
\end{equation}
(with the trace extending over the Hilbert space) gives rise to 
periodic boundary conditions in the extra dimension.

The ground state of the quantum spin system has a broken global 
$SU(N)$ symmetry. The choice of the $SU(N)$ representation determines the
symmetry breaking pattern. We choose a totally symmetric $SU(N)$ representation
corresponding to a Young tableau with a single row containing $n$ boxes. It is 
easy to construct the ground states of the $SU(N)$ ferromagnet, and one finds 
spontaneous symmetry breaking from $SU(N)$ to $U(N-1)$. Consequently, there are
$(N^2 - 1) - (N-1)^2 = 2N - 2$ massless Goldstone bosons described by fields 
$P(x)$ in the coset space $SU(N)/U(N-1) = CP(N-1)$. In the leading order of 
chiral perturbation theory the Euclidean action for the Goldstone boson fields 
is given by
\begin{equation}
\label{ferroaction}
S[P] = \int_0^\beta dt \int d^2x \ \mbox{Tr}
[\rho_s \partial_\mu P \partial_\mu P - \frac{2 n}{a^2} \int_0^1 d\tau \ 
P \partial_t P \partial_\tau P].
\end{equation}
Here $\rho_s$ is the spin stiffness, which is analogous to the pion decay
constant in QCD. The second term in eq.(\ref{ferroaction}) is a 
Wess-Zumino-Witten 
term which involves an integral over an interpolation parameter
$\tau$.

For $\beta = \infty$ the system then
has a spontaneously broken global symmetry and thus massless Goldstone bosons.
However, as soon as $\beta$ becomes finite, due to the 
Hohenberg-Mermin-Wagner-Coleman theorem, the symmetry can no longer be broken,
and, consequently, the Goldstone bosons pick up a small mass $m$
nonperturbatively. As a result, the corresponding correlation length 
$\xi = 1/m$ becomes finite and the $SU(N)$ symmetry is restored over that 
length scale. The question arises if $\xi$ is bigger or smaller than the
extent $\beta$ of the extra dimension. When $\xi \gg \beta$ the Goldstone boson
field is essentially constant along the extra dimension and the system 
undergoes dimensional reduction. Since the Wess-Zumino-Witten term
vanishes for field constant in $t$, after dimensional reduction the
action reduces to
\begin{equation}
\label{targetaction}
S[P] = \beta \rho_s \int d^2x \ \mbox{Tr}[\partial_\mu P \partial_\mu P],
\end{equation}
which is just the action of the 2-d target $CP(N-1)$ model. The coupling
constant of the 2-d model is determined by the extent of the extra dimension
and is given by
\begin{equation}
\frac{1}{g^2} = \beta \rho_s.
\end{equation}
Due to asymptotic freedom of the 2-d $CP(N-1)$ model, for small $g^2$ the 
correlation length is exponentially large, i.e.\
\begin{equation}
\xi \propto \exp(4 \pi \beta \rho_s/N).
\end{equation}
Here $N/4 \pi$ is the 1-loop coefficient of the perturbative $\beta$-function.
Indeed, one sees that $\xi \gg \beta$ as long as $\beta$ itself is sufficiently
large. In particular, somewhat counter-intuitively, dimensional reduction 
happens in the large $\beta$ limit because $\xi$ then grows exponentially. In
D-theory one approaches the continuum limit not by varying a bare coupling 
constant but by increasing the extent $\beta$ of the extra dimension. This
mechanism of dimensional reduction of discrete variables is generic and occurs
in all asymptotically free D-theory models \cite{Bro99,Bro04}. It should
be noted that (just like in the standard approach) no fine-tuning is needed
to approach the continuum limit.

\section{Path Integral Representation of $SU(N)$ Quantum Spin Systems}

Let us construct a path integral representation for the partition function $Z$
of the $SU(N)$ quantum spin ferromagnet introduced above. In an intermediate 
step we introduce a lattice in the 
Euclidean time direction, using a Trotter decomposition of the Hamiltonian.
However, since we are dealing with discrete variables, the path integral is
completely well-defined even in continuous Euclidean time. Also the cluster 
algorithm to be described in the following section can operate directly in the
Euclidean time continuum \cite{Beard96}. Hence, the final results are
completely independent
of the Trotter decomposition. In $2$ spatial dimensions (with an even extent) 
we decompose the Hamilton operator into $4$ terms
\begin{equation}
H = H_1 + H_2 + H_3 + H_4,
\end{equation}
with
\begin{equation}
H_{1,2} = \!\! \sum_{\stackrel{x = (x_1,x_2)}{x_i \rm{even}}} \!\! 
h_{x,i}, \ \
H_{3,4} = \!\! \sum_{\stackrel{x = (x_1,x_2)}{x_i \rm{odd}}} \!\! 
h_{x,i}.
\end{equation}
The individual contributions
\begin{equation}
h_{x,i} = - J \ T_x^a T_{x+\hat i}^a,
\end{equation}
to a given $H_i$ commute with each other, but two different $H_i$ do not 
commute. Using the Trotter formula, the partition function then takes the form
\begin{eqnarray}
Z\!\!\!&=&\!\!\!\lim_{M \rightarrow \infty} \! \mbox{Tr} 
\left\{\exp(- \epsilon H_1) \exp(- \epsilon H_2) \exp(- \epsilon H_3)
  \exp(- \epsilon H_4) \right\}^M.
\end{eqnarray}
We have introduced $M$ Euclidean time-slices with $\epsilon = \beta/M$ being 
the lattice spacing in the Euclidean time direction. Inserting complete sets of spin states $q \in \{u,d,s,...\}$ the partition 
function takes the form
\begin{equation}
Z = \sum_{[q]} \exp(- S[q]).
\end{equation}
The sum extends over configurations $[q]$ of spins $q(x,t)$ on a 
$(2+1)$-dimensional space-time lattice of points $(x,t)$. The Boltzmann factor
is a product of space-time plaquette contributions with
\begin{eqnarray}
\label{Boltzmannf}
&&\exp(- s[u,u,u,u]) = \exp(- s[d,d,d,d]) = 1,
\nonumber \\
&&\exp(- s[u,d,u,d]) = \exp(- s[d,u,d,u]) = 
\frac{1}{2}[1 + \exp(- \epsilon J)],
\nonumber \\
&&\exp(- s[u,d,d,u]) = \exp(- s[d,u,u,d]) =
\frac{1}{2}[1 - \exp(- \epsilon J)].
\end{eqnarray}
In these expressions the flavors $u$ and $d$ can be permuted to other values. 
All the other Boltzmann factors are zero, which implies several constraints on 
allowed configurations. 

\section{Cluster Algorithm for $SU(N)$ Quantum Ferromagnets}

Let us now discuss the cluster algorithm for the $SU(N)$ quantum ferromagnet.
Just like the original $SU(2)$ loop-cluster algorithm \cite{Eve93,Wie94}, the
$SU(N)$ cluster algorithm builds a closed loop connecting neighboring lattice 
points with the spin in the same quantum state, and then changes the state of
all those spins to a different randomly chosen common value. To begin cluster 
growth, an initial lattice point $(x,t)$ is picked at random. The spin located
at that point participates in two plaquette interactions, one before and one 
after $t$. One picks one interaction arbitrarily and considers the states
of the other spins on that plaquette. One of the corners of this interaction
plaquette will be the next point on the loop. For configurations
$C_1 = [u,d,u,d]$ or $[d,u,d,u]$ the next point is the time-like neighbor of
$(x,t)$ on the plaquette, while for configurations $C_2 = [u,d,d,u]$ or 
$[d,u,u,d]$ the next point is the diagonal neighbor. If the states are all the
same, i.e.\ for $C_3 = [u,u,u,u]$ or $[d,d,d,d]$, with probability 
\begin{equation}
p = \frac{1}{2}[1 + \exp(-\epsilon J)]
\end{equation}
the next point on the loop is again the time-like neighbor, and with 
probability $(1 - p)$ it is the diagonal neighbor. The next point on the loop
belongs to another interaction plaquette on which the same process is 
repeated. In this way the loop grows until it finally closes.

\section{Critical slowing down in the continuum limit}

In order to determine the efficiency of this algorithm one has to study
its critical slowing
down  when one approaches the continuum limit. We have
used a multi-cluster algorithm for an $SU(3)$ quantum
ferromagnet which corresponds to a $CP(2)$ model. As an observable, we
have chosen the uniform magnetization which gives the cleanest signal,
\begin{equation}
M= \sum_{x,t} (\delta_{q(x,t),u}-\delta_{q(x,t),d}).
\end{equation}
The autocorrelation time $\tau$
of the magnetization is determined from the exponential fall-off of the
autocorrelation function. The simulations have been performed  at fixed
ratio $\xi/\/L\approx2.5$, for lattice sizes
$L/a$ = 20, 40, 80, 160, 320, 640 and the corresponding correlation lengths
$\xi/a$ = 8.87(1), 16.76(1), 32.26(3), 64.6(1), 123.4(2),
253(1). Remarkably, the autocorrelation time doesn't change
when one varies the size of the system and the correlation length and
stays close to
$\tau\approx1$ sweep. This is a strong indication for an almost
perfect algorithm where the critical slowing down is completely
eliminated.

\section{Conclusions}

Due to a no-go theorem \cite{Caracciolo93}, so far no efficient
cluster algorithm has been
developed for $CP(N-1)$ models in the usual Wilson formulation. In the
D-theory formulation, one has been able to perform simulations using a
multi-cluster algorithm for large
correlation lengths and with a corresponding autocorrelation time of about
one sweep. Remarkably, there is almost no variation of the
autocorrelation time when one spans a factor of about 30 in the correlation
length. The critical slowing down of the algorithm in the continuum
limit is hence completely eliminated. These results can be
compared to the ones obtained with the efficient multigrid
algorithm \cite{has93}. Our method has the advantage to obtain
autocorrelation times more than 20 times smaller for similar
correlation lengths, the multi-cluster algorithm is in addition
straightforward to implement. \\ 
With the D-theory regularization, it has recently also been possible to
simulate $CP(N-1)$ models at non-trivial $\theta$-vacuum angle
\cite{bbb04} which is normally impossible due to a severe sign
problem.


\begin{thebibliography}{99}

\bibitem{DAd78}
A.\ D'Adda, P.\ Di Vecchia, and M.\ L\"uscher, \emph{An $1/N$
  expandable series of nonlinear sigma model with instantons}, Nucl.\
Phys.\ {\bf B146} (1978) 63. 

\bibitem{Bro99}
R.\ Brower, S.\ Chandrasekharan, and U.-J.\ Wiese, \emph{$QCD$ as a
  quantum link model}, Phys.\ Rev.\ {\bf D60} (1999) 094502 [{\tt
  hep-th/9704106}]. 


\bibitem{Bro04}
R.\ Brower, S.\ Chandrasekharan, S.\ Riederer, and U.-J.\ Wiese,
\emph{D-Theory: Field quantization by dimensional reduction of
  discrete variables}, Nucl.\ Phys.\ {\bf B693} (2004) 149 [{\tt hep-lat/0309182}].


\bibitem{Beard96}
B. B. Beard, U.-J. Wiese, \emph{Simulations of discrete quantum
  systems in continuous Euclidean time}, Phys. Rev. Lett. {\bf 77} (1996)
5130 [{\tt cond-mat/9602164}].


\bibitem{Eve93}
H.\ G.\ Evertz, G.\ Lana, and M.\ Marcu,  \emph{Cluster algorithm for
  vertex models}, Phys. Rev. Lett. {\bf 70} (1993) 875-879 [{\tt cond-mat/9211006}].


\bibitem{Wie94}
U.-J.\ Wiese and H.-P.\ Ying, \emph{A determination of the low-energy
  parameters of the 2-d Heisenberg antiferromagnet}, Z. Phys. {\bf B93}
(1994) 147 [{\tt cond-mat/9212006}].


\bibitem{Caracciolo93}
S. Caracciolo, R. G. Edwards, A. Pelissetto, A. D. Sokal, \emph{Wolff
  type embedding algorithm for general nonlinear sigma models},
Nucl. Phys. {\bf B403} (1993) 475-541 [{\tt hep-lat/9205005}].


\bibitem{has93}
M. Hasenbusch and S. Meyer, \emph{Testing accelerated algorithms in the
  lattice $CP(3)$ model}, Phys. Rev. {\bf D45} (1992) 4376-4380.

\bibitem{bbb04}
B.B. Beard, M. Pepe, S. Riederer, U.-J. Wiese, \emph{Study of
  $CP(N-1)$ theta-vacua by cluster-simulation of $SU(N)$ quantum spin
  ladders}, Phys. Rev. Lett. {\bf 94} (2005) 010603 [{\tt hep-lat/0406040}].


\end{thebibliography}
\end{document}